\documentclass[showpacs,amsmath,amssymb,nofootinbib]{revtex4}
\usepackage{color}
\usepackage{epsfig}
\begin{document}
\title{Hadron collider potential for excited bosons search: A Snowmass
whitepaper}
\author{M.V. Chizhov$^{1,2}$, V. A. Bednyakov$^1$, J. A. Budagov$^1$}
\affiliation{$^{\it 1}$Dzhelepov Laboratory of Nuclear Problems,\\
\mbox{Joint Institute for Nuclear Research, 141980, Dubna,
Russia}\\
$^{\it 2}$Centre for Space Research and Technologies, Faculty of
Physics, Sofia University, 1164 Sofia, Bulgaria}

\begin{abstract}
The dilepton final states, $e^+e^-$ and $\mu^+\mu^-$, are the most
clear channels for new heavy neutral resonances search. Their
advantage is that the main irreducible background from the Standard
Model Drell--Yan process contributes usually two orders of magnitude
lower than the expected signal under the peak region. In this paper
we are focused on the search of the excited neutral bosons
$Z^\star$. At present only the ATLAS Collaboration is looking for
such excitations at LHC. We compare our evaluations with the
official collaboration results at 7~TeV, and present our estimations
at higher centre-of-mass energies in $pp$ collisions and different
luminosities.
\end{abstract}

 \maketitle

\section{Introduction}

The idea of compositeness of the nature is not new. However, in
order to explore the internal structure of the matter, the
high-energy colliders are necessary. That is why, for example,
search of hypothetical {\em excited fermions}
${\color{red}\psi^\star}$ has been fulfilled at all powerful
colliders, such as LEP~\cite{LEP}, HERA~\cite{HERA},
Tevatron~\cite{Tevatron} and continues at
LHC~\cite{LHCleptons,LHCdijet}.

The excited fermions have anomalous (magnetic moment type) couplings
with the known fermions $\psi$ and the gauge bosons (such as gluons,
photons and weak $W/Z$ bosons)
\begin{equation}\label{psi}
    {\cal L}_{\rm excited}^{\color{red}\psi^\star}=
    \frac{g}{2\Lambda}\:{\color{red}\bar{\psi}^\star}\sigma^{\mu\nu}\!\psi\:
    \left(\partial_\mu Z_\nu-\partial_\nu Z_\mu\right)
    +{\rm h.c.},
\end{equation}
where the parameter $\Lambda$ is connected to the compositeness mass
scale of the new physics. Due to their anomalous type of couplings
they lead to a unique experimental signature for their detection.

The interaction (\ref{psi}) could be also reinterpreted from a
different point of view, introducing the new {\em excited bosons}
${\color{red}Z^\star}$~\cite{project}
\begin{equation}\label{Z*}
    {\cal L}_{\rm excited}^{\color{red}Z^\star}=
    \frac{g}{2\Lambda}\:\bar{\psi}\,\sigma^{\mu\nu}\!\psi\:
    \left(\partial_\mu{\color{red} Z^\star_\nu}
    -\partial_\nu{\color{red} Z^\star_\mu}\right)
\end{equation}
instead of the fermionic ones. Such type of new heavy bosons
${\color{red}Z^\star}$ could also be interesting objects for
experimental searches due to their different couplings to the
ordinary fermions in comparison with the minimal gauge
${\color{blue}Z'}$ couplings. In this paper we are focused on the
search of the excited bosons ${\color{red}Z^\star}$, rather than on
the well-known ${\color{blue}Z^\prime}$ bosons from various
benchmark models.

In contrast with the minimal gauge couplings, where either only
left-handed or right-handed fermions participate in the
interactions, the tensor currents mix both left-handed and
right-handed fermions. Therefore, like the Higgs particles, the
excited bosons carry a nonzero chiral charge and according to the
symmetry of the Standard Model they should be introduced as the
electroweak doublets
$({\color{red}Z^\star\;W^\star})$~\cite{doublets} with the internal
quantum numbers identical to the Standard Model Higgs doublet.

The existence of such doublets with masses not far from the weak
scale is motivated by the hierarchy problem~\cite{CD}. The effective
interaction (\ref{Z*}) is induced by quantum loop corrections from a
renormalizable underlying theory and represents the lowest order
effective Lagrangian for the excited bosons interacting with the
Standard Model fermions. The corresponding reference model is
described in \cite{ref}.

Compared to other heavy bosons, interactions mediated by
$({\color{red}Z^\star\;W^\star})$ doublets are additionally
suppressed in low-energy processes by powers of small ratio of the
momentum transfer to the parameter $\Lambda$. Thus, the search of
the excited bosons is especially motivated at the LHC and future
colliders and at present is conducted by the ATLAS
Collaboration~\cite{Zstar,Wstar}. Besides this, the derivative
couplings lead to unique signatures for detection of such bosons at
the hadron colliders. Decay products of the excited bosons possess
previously unexplored angular distribution, which leads to a new
strategy of the resonance search and identification in
dilepton~\cite{dilepton} and dijet~\cite{dijet} channels.

\begin{figure}[h!]\centering
\epsfig{file=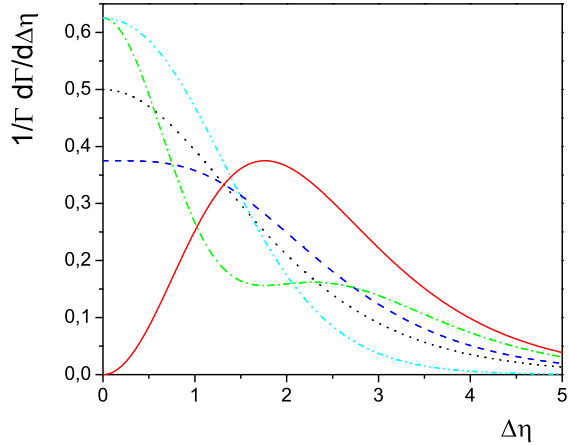,width=0.5\textwidth} \caption{\label{fig:spin}
The normalized angular final fermions distributions as functions of
$\Delta\eta$ for the scalar (dotted), spin-1 bosons with the minimal
couplings (dashed), the excited bosons (solid) and spin-2
resonances, produced through quark (dash-dotted) and gluon
(dash--double-dotted) fusion, are shown.}
\end{figure}

The crucial variable, which can help to differ the decay
distribution from other resonances, is an absolute value of the
pseudorapidity difference $\Delta\eta\equiv|\eta_1-\eta_2|$ between
the final fermions (see Fig.~\ref{fig:spin}).
Decay distributions of all other resonances have the kinematic
absolute maximum at $\Delta\eta=0$, while the excited bosons decay
distribution is zero at this point and peaks at
$\Delta\eta=\ln(3+\sqrt{8})\approx 1.76$. The latter corresponds to
the polar angle $\theta = 45^\circ$ for the final fermions in the
resonance rest frame and a little bit contradicts to the common
opinion about an expected signal from new physics at $\theta =
90^\circ$.

The background from the Standard Model Drell--Yan (DY) process
contributes mainly to the central pseudorapidity region
$\Delta\eta\approx0$ from the intermediate $\gamma/Z$ bosons, which
have the minimal gauge couplings with quarks and leptons. The
background can be suppressed up to 40\% with appropriate cut
$\Delta\eta > \Delta\eta_{\rm min}\approx 1.0$ leaving the main part
of the signal intact. This allows to enhance the significance of
bump search for the excited bosons in the dilepton
channels~\cite{dilepton}.

However, for dijet final states the huge QCD background is
exponentially dominated 
at high $\Delta\eta$ due to $t$-channel gluon exchanges, which
possess a Rutherford-like distribution $1/(1-\cos \theta)^2$. It is
the reason ATLAS and~CMS Collaborations to apply severe cut from
above $\Delta\eta < \Delta\eta_{\rm max}\approx
1.2\div1.3$~\cite{LHCdijet}. Such low value of $\Delta\eta_{\rm
max}$ is optimal for resonance searches with nearly isotropic decay
distributions, but is not optimal for the excited bosons, where the
most of the signal is removed. Therefore, in order to optimize
signal significance for the excited bosons the corresponding cut
should be elevated~\cite{dijet} even allowing more background
events.

In this paper we investigate hadron collider potential for excited
bosons search in the most clear (dilepton) channels. Usually, using
these channels leads to more severe constraints on resonance mass
than from dijets channels. We compare our evaluations with the
official ATLAS Collaboration results at 7~TeV~\cite{Zstar}, and
present our estimations at higher center-of-mass energies in $pp$
collisions and different luminosities.

\section{Signal and background samples}

In order to generate signal and background samples we use the
CalcHEP package~\cite{CalcHEP}. Although the package allows to
perform calculations only in Born (LO) approximation, using the same
generator provides some uniformity between signal and background
generation. With its batch and web-interface facilities the CalcHEP
has become user-friendly program. Besides this, High Energy Physics
Model DataBase (HEPMDB) system~\cite{hepmdb} and IRIDIS High
Performance Computing cluster at the University of Southampton
provide access to different theoretical models and fast computer
nodes. The authors acknowledge the use of these facilities in the
completion of this work.

We have used the simplified reference model ESM~\cite{ref} for the
excited bosons. The only ``down-type'' neutral
${\color{red}Z^\star}$ boson interacts both with quarks and charged
leptons. Therefore, it can be produced at hadron colliders and can
be seen in the leptonic final states in the DY process. To
investigate the resonance shape in the invariant dilepton mass and
other distributions for various resonance pole masses of
${\color{red}Z^\star}$ bosons many signal samples should be
generated. In this paper we apply a template
technique~\cite{template} both for signal and background samples.
It allows to generate only one sample both for signal and
background. The necessary distributions for the fixed pole mass $M$
can be obtained by reweighting the corresponding samples.

\begin{figure}[h!]\centering
\epsfig{file=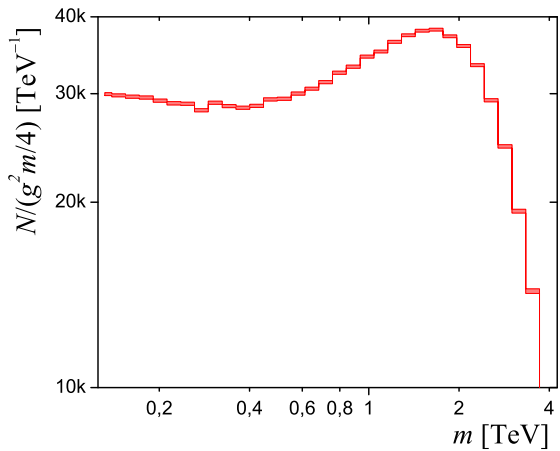,width=0.5\textwidth}
\caption{\label{fig:template} $Z^\star$ template distribution at
$\sqrt{s}=8$ TeV.}
\end{figure}

The signal template sample is generated without Breit--Wigner pole
mass factor
\begin{equation}\label{BW}
    BW(m)=\frac{1}{(m^2-M^2)^2+(\Gamma M)^2},
\end{equation}
and with correction function
\begin{equation}\label{f}
    f(m)=m^\alpha\exp(\beta m)
\end{equation}
of luminosity decreasing at higher dilepton invariant masses $m$. It
can be realized using CalcHEP user function \verb"usrFF.c". The
latter correction (\ref{f}) is fulfilled to ensure the same relative
errors in reweighted samples regardless of the resonance pole mass.
This can be achieved by choosing the constants $\alpha$ and $\beta$
in such a way that the resulting template distribution decreases
inversely with the invariant mass, namely $C/m$. Therefore, it
should be flat in logarithmic invariant mass scale (see
Fig.~\ref{fig:template}).

Since the total resonance decay width into fermions
\begin{equation}\label{gamma}
    \Gamma = \frac{g^2}{4\pi}M\approx 0.034 M
\end{equation}
is proportional to the resonance mass, the dimensionless variable
$x=m/M$ can be introduced in (\ref{BW}). Then the total number of
events and the absolute error after reweighting can be estimated in
the resonance vicinity $x\sim 1$ as
\begin{eqnarray}\label{events}
    N_{\rm rew}&\simeq&\frac{f^{-1}(M)}{M^4}\int_{0}^{\infty}
    \frac{C{\rm d}x}{(x^2-1)^2+(g^2/4\pi)^2},\\
\label{de}
    {\Delta}N_{\rm rew}&\simeq&\frac{f^{-1}(M)}{M^4}\,\sqrt{\int_{0}^{\infty}
    \frac{C{\rm d}x}{[(x^2-1)^2+(g^2/4\pi)^2]^2}}.
\end{eqnarray}
It is valid only for narrow resonances and infinite detector
resolution. From eqs. (\ref{events}) and (\ref{de}) it is clear that
the relative error in the reweighted sample will be approximately
the same for any resonance mass from the range $[M_{\rm min},M_{\rm
max}]$ and proportional to
\begin{equation}\label{re}
    \frac{{\Delta}N_{\rm rew}}{N_{\rm rew}}\simeq\frac{2}{g}
    \sqrt{\frac{\ln(M_{\rm max}/M_{\rm min})}{N}},
\end{equation}
where $N$ is the total number of generated events in the template
sample.

\begin{figure}[h!]\centering
\epsfig{file=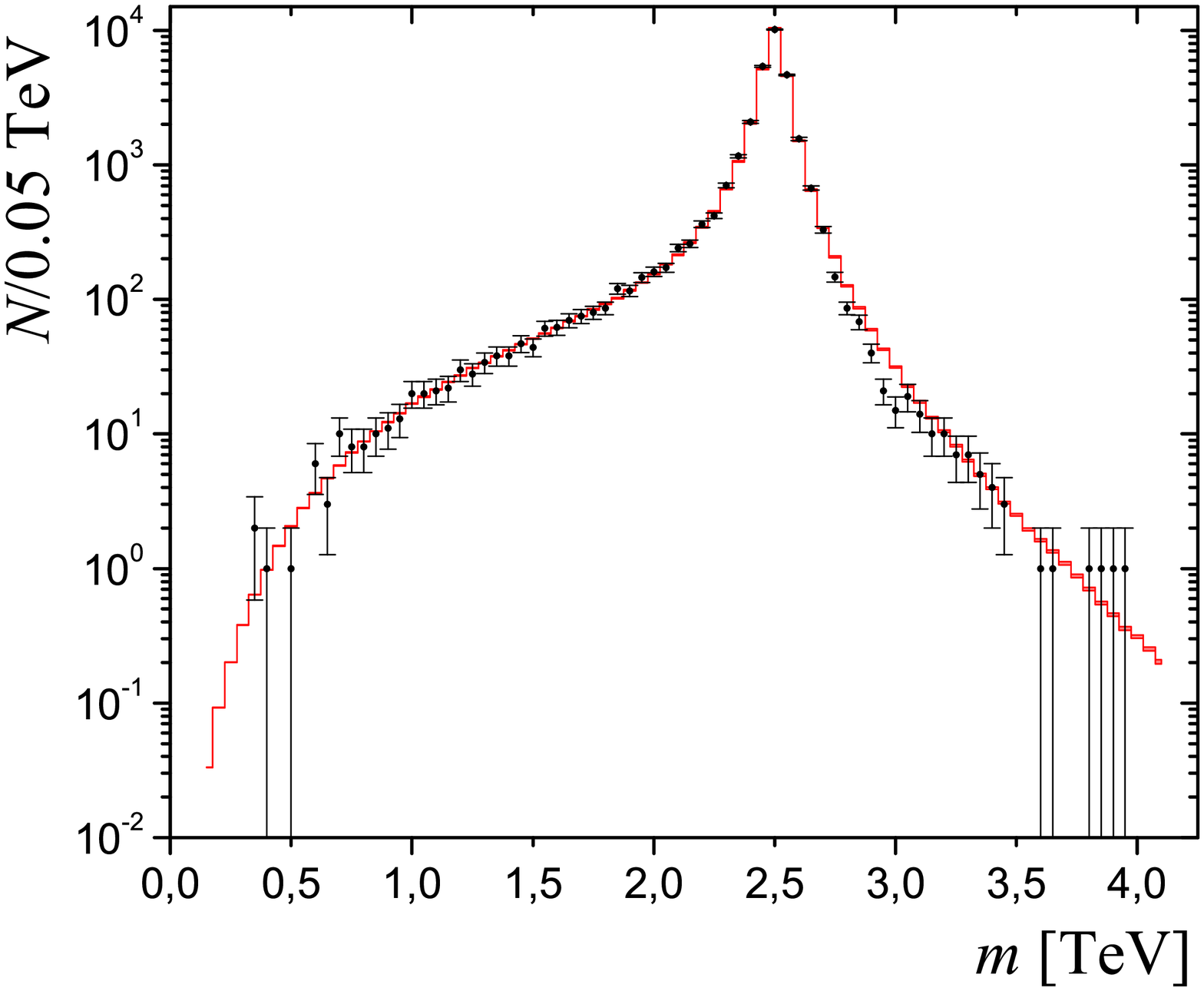,width=0.49\textwidth}
\epsfig{file=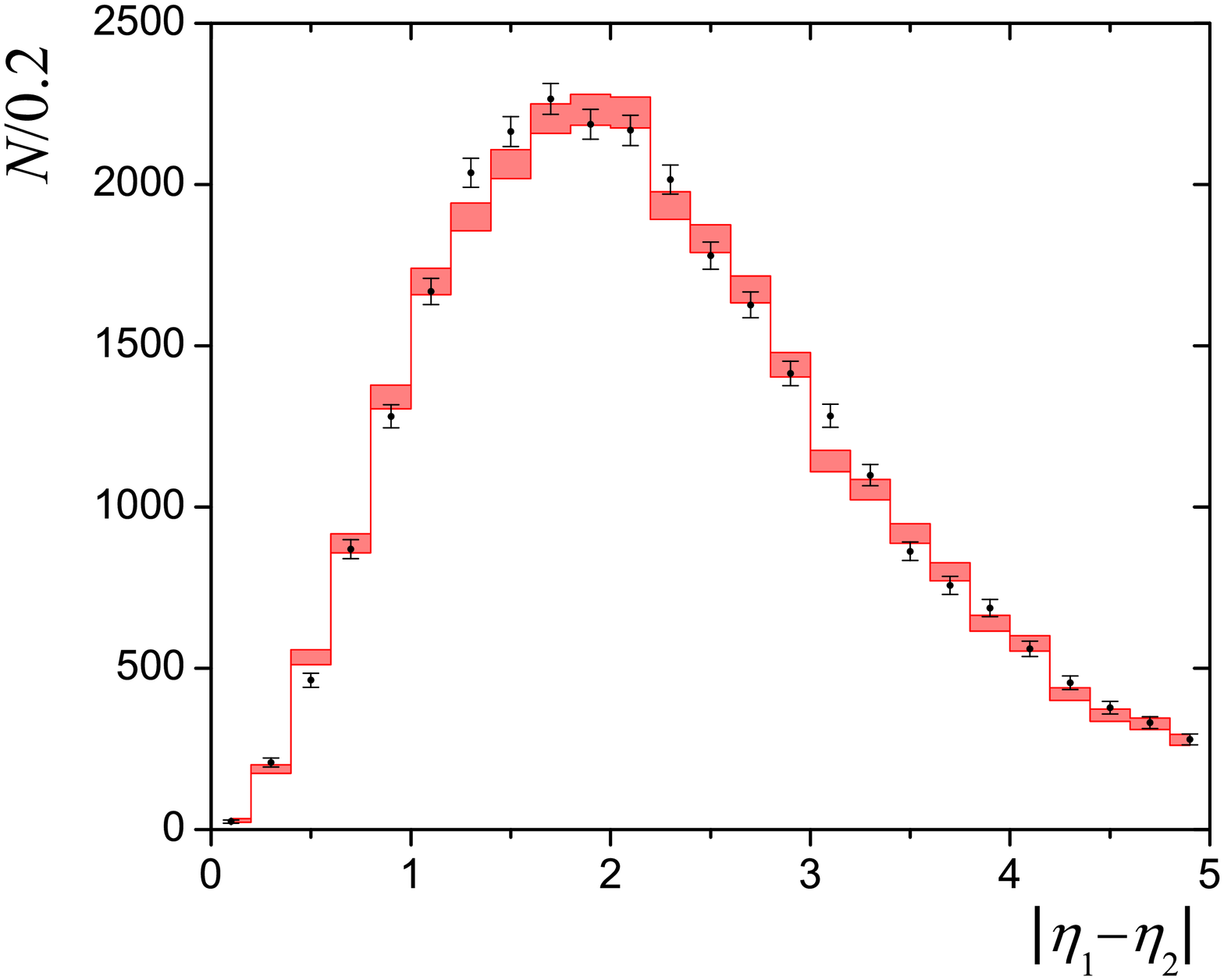,width=0.49\textwidth}
\caption{\label{fig:reweight} Comparison between the invariant mass
(left) and the pseudorapidity difference (right) distributions for
dedicated sample of 2.5 TeV pole mass resonance (points with error
bars) and reweighted template sample (histograms) at $\sqrt{s}=8$
TeV is shown.}
\end{figure}

In Fig.~\ref{fig:template} the {\it LogFlat} signal template
distribution at $\sqrt{s}=8$ TeV is shown, which contains 1 Mevents.
The special bin size $g^2m/4\approx 0.1 m$ is selected to show the
number of events, which dedicated sample for a fixed pole mass
should contain in order to achieve a comparable precision with the
reweighted sample from the given template (see eq.~(\ref{re})). To
validate this statement a comparison between distributions for
dedicated sample of 2.5~TeV pole mass resonance produced at
$\sqrt{s}=8$ TeV containing 30 kevents and reweighted template
sample is shown in Fig.~\ref{fig:reweight}.

To derive the exclusion limits and the discovery potential we should
compare the signal and the background. In this paper the simplest
``number counting'' approach is adopted, which is based on the
comparison of the expected rate of events for the signal and the
background processes. From these rates, and assuming Poisson
statistics, one can determine the probability that background
fluctuations produce a signal-like result according to some
estimator, e.g. the likelihood ratio.

\begin{figure}[h!]\centering
\epsfig{file=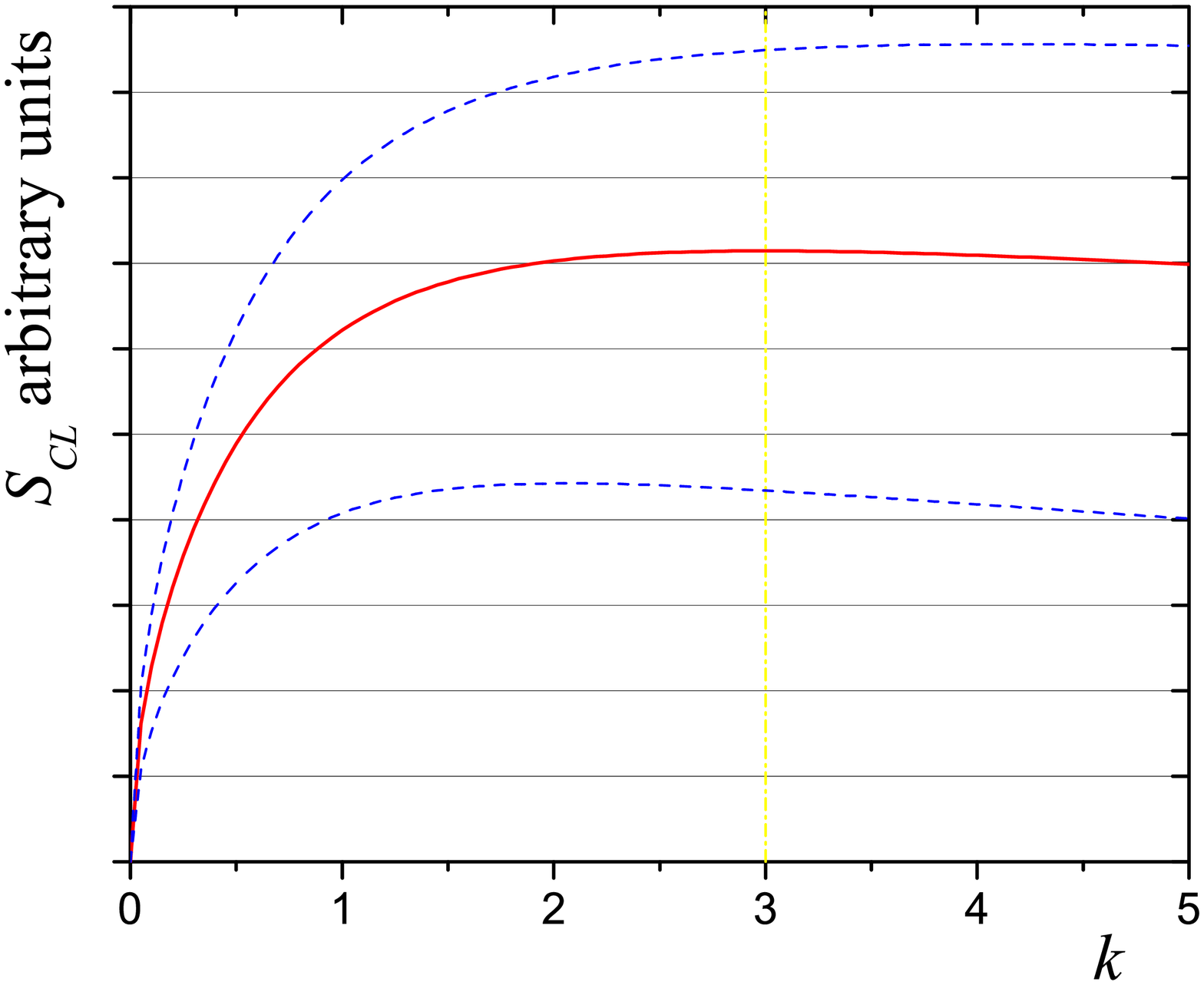,width=0.4\textwidth} \caption{\label{fig:Scl}
The signal significance curves for different signal to background
ratios as a function of the window size.}
\end{figure}

For the narrow resonances the bulk of events populates the vicinity
around the peak $[M-k\Gamma,M+k\Gamma]$. The relative ratio of the
signal events in this region to the total events number can be
estimated using eq.~(\ref{BW}) as
\begin{equation}\label{rpeak}
    s = \frac{2\arctan(k)}{\pi}.
\end{equation}
If we assume that the background contribution is proportional to the
size of the on-peak region $b \sim k$, we can estimate analytically
the maximum of the signal significance on $k$ using the formula from
Appendix A of Ref.~\cite{CMS}
\begin{equation}\label{Scl}
    S_{CL}=\sqrt{2\left((s+b)\ln\left(1+\frac{s}{b}\right)-s\right)},
\end{equation}
which follows directly from the Poisson distribution. We will use
this equation for estimation of discovery potential. The
corresponding curves as a function of the window size are shown in
Fig.~\ref{fig:Scl}.
The middle curve reaches maximum at $k\simeq 3$ for $s/b\simeq
43.8$, which corresponds, for example, to resonance production with
$M=3.25$~TeV at $\sqrt{s}=8$~TeV. The upper and bottom curves
correspond to 10 times bigger and lower signal to background ratios,
respectively. They reach maxima at a little bit bigger and lower
size windows than the middle curve, correspondingly. It is in a good
agreement with the numerical calculations (see Fig.~2 from
\cite{Lathule}). In the following we will use the optimal average
value $k=3$.

However, in the muon channel the experimental $p_{\rm T}$ resolution
increases as $p^2_{\rm T}$ at high muon momenta. It means that at
some point we cannot neglect anymore the experimental resolution in
comparison with resonance width, which increases linearly with
resonance mass (see eq.~(\ref{gamma})). Therefore, in this case we
cannot choose the optimal window due to experimental resolution and
oblige to increase it allowing more background, that directly
affects the discovery potential and the exclusion limits in the muon
and combined channels. We are thankful to Igor Boyko for this
remark.

\section{Exclusion limits at $\sqrt{s}=7$~TeV and comparison with ATLAS results.}

At present only the ATLAS Collaboration is looking for a production
of the excited ${\color{red}Z^\star}$ bosons \cite{Zstar10,Zstar}.
In 2011 the ATLAS Collaboration collected 4.9 fb$^{-1}$ of an
integrated luminosity in dielectron channel and 5.0 fb$^{-1}$ in
dimuon channel at a center-of-mass energy of 7 TeV. The 95\% CL
observed and the expected exclusion limits are shown in Table~I
\cite{Zstar}.
\begin{table}[h]
  \label{tab:limits7}
\begin{center}
\begin{tabular}{|c|ccc|}
  \hline
   & $Z^\star\to \mu^+\mu^-$ & $Z^\star\to e^+e^-$ & $Z^\star\to \ell^+\ell^-$ \\
  \hline
  Observed limit [TeV] & 1.97 & 2.10 & 2.20 \\
  Expected limit [TeV] & 1.99 & 2.13 & 2.22 \\
  \hline
\end{tabular}
  \caption{The observed and expected 95\% CL lower limits on the mass of the $Z^\star$
  boson for the $\mu^+\mu^-$ and $e^+e^-$ channels separately and for their
  combination from \cite{Zstar}.}
\end{center}
\end{table}

\noindent The combination of the dielectron and dimuon channels is
performed under the assumption of lepton universality. The combined
limit on the cross section times branching fraction ($\sigma Br$)
expected from theory is around 0.7~fb and can be read from Fig.~4 of
\cite{Zstar}. Since there are no events above 2~TeV dilepton
invariant mass and the background is very small, the observed and
expected limits are nearly the same.

The ATLAS Collaboration has used the Bayesian approach \cite{Bayes}
with a flat, positive prior on the signal cross section to determine
an upper limit on the number of signal events. In this paper we will
use more simple ``number counting'' approach for the exclusion limit
and the discovery potential evaluations with many other
approximations. Nevertheless, we will show that our limit
estimations are in agreement with the official ATLAS results.

Since there is not yet a deviation from the Standard Model
distributions only the exclusion limits can be evaluated. To do this
we will use an approximate computation of the confidence level for
combining searches with small statistics~\cite{Junk}. The
$[M-k\Gamma,M+k\Gamma]$ on-peak region with $k=3$ has been used for
event counting in both channels for the resonance masses below
2~TeV. However, due to bad experimental resolution in the muon
channel for the resonance masses above 2~TeV we assume
$k=3M{\rm[TeV]}/2{\rm~TeV}$. We will take into an account only the
leading $Z/\gamma$ DY background and neglect many subdominant
backgrounds, like QCD, $t\bar{t}$, diboson and $W+$~jets.

The number of the expected events, recorded in the detector and that
have passed a selection criteria, depends on many factors, which can
be expressed through the overall event acceptance times efficiency
($\cal{A}\epsilon$). So, for a ${\color{blue}Z^\prime}$ boson of
mass 2~TeV decaying into a dielectron final state the overall event
$\cal{A}\epsilon$ is about 66\%, while for the muon channel this
factor is only 43\% for the ATLAS detector during 2010 data taking
period \cite{Zstar10}. To be specific, we will accept the following
rounded overall event $\cal{A}\epsilon$ numbers for the ATLAS
detector: 70\% for the electron channel and 40\% for the muon
channel, both for signal and background.

In order to follow closely as possible the ATLAS analysis in this
section we will use  the same PDF set, CTEQ6L1~\cite{cteq6l1}, used
by ATLAS for signal generation. Since only the central value is
available for the CTEQ6L1 PDF set, the closest set,
CTEQ61~\cite{cteq61}, is used to estimate systematic PDF
uncertainties (see Table~II).
\begin{table}[h]
  \label{tab:pdf7}
\begin{center}
\begin{tabular}{||r|cc|cc||r|cc|cc||}\hline
      $Z^*$ mass &  \multicolumn{2}{|c|}{signal} &
      \multicolumn{2}{|c||}{background} &$Z^*$ mass &
      \multicolumn{2}{|c|}{signal} & \multicolumn{2}{|c||}{background} \\
      $$[GeV]   &   $\sigma Br$ [fb] & $\Delta\sigma/\sigma$ [\%]
      & $\sigma Br$ [fb] & $\Delta\sigma/\sigma$ [\%] &
      $$[TeV] & $\sigma Br$ [ab] & $\Delta\sigma/\sigma$ [\%]
      & $\sigma Br$ [ab] & $\Delta\sigma/\sigma$ [\%] \\ \hline
       250  &  64866.  & 5.1 & 355.66 & 4.4 & 2.00 & 1565.2 & 32.1 & 15.248 & 17.5\\
       500  &  4608.2   & 7.0 & 25.418 & 5.4 & 2.25 & 507.93 & 41.9 & 5.8910 & 22.1\\
       750   &  779.58   & 9.5 & 4.5352 & 6.4 & 2.50 & 165.43 & 54.8 & 2.3148 & 27.7\\
       1000  &  184.29   & 11.9 & 1.1399 & 7.4 & 2.75 & 53.988 & 72.0 & 0.91438 & 34.5\\
       1250  &  51.121   & 15.4 & 0.34161 & 8.4 & 3.00 & 17.720 & 93.3 & 0.35945 & 42.8\\
       1500  &  15.436   & 19.3 & 0.11385 & 10.7 & 3.25 & 5.8534 & 118.4& 0.13933 & 53.1\\
       1750  &  4.8631   & 24.9 & 0.040702 & 13.6 & 3.50 & 1.9359 & 146.6 & 0.052909 & 65.8\\ \hline
\end{tabular}  \caption{ $Z^\star$ signal and the Standard Model DY background
cross sections times branching fraction in $[M-3\Gamma,M+3\Gamma]$
on-peak region and maximal relative uncertainty due to PDF variation
(at 90\% C.L.) with CTEQ61 set.}
\end{center}
\end{table}

The PDF uncertainties dominate at high dilepton invariant masses.
Therefore, we will use only them as systematic uncertainties. Using
program \verb"eclsyst.f" \cite{Junk} we get the following limits on
$Z^\star$ mass: 2.01~TeV in the muon channel and 2.15~TeV in the
electron channel. Their combination excludes at 95\% confidence
level $Z^\star$ masses below 2.25~TeV and $\sigma Br>0.7$~fb. The
obtained results are very close to the official ATLAS results and
convince us to investigate the LHC potential at higher
center-of-mass energies.

\section{Exclusion limits at $\sqrt{s}=8$~TeV.}

In 2012 the ATLAS experiment recorded 20~fb$^{-1}$ of good data both
in electron and muon channels at $\sqrt{s}=8$~TeV. Again good
agreement between the data and the background expectation was found
\cite{ATLAS8}. However, limits only on the $Z^\prime$ Sequential
Standard Model boson, $E_6$ gauge bosons and a spin-2
Randall-Sundrum graviton have been set so far. There are still no
events observed by ATLAS~\cite{ATLAS8} above 2~TeV dilepton
invariant mass and the observed and expected limits are nearly the
same.

In this paper we precede the official ATLAS results on the excited
boson search and evaluate exclusion limits on ${\color{red}Z^\star}$
at $\sqrt{s}=8$~TeV. The analysis is fulfilled in the same lines as
in the previous section. The only difference, that we will use more
recent MSTW2008lo PDF set \cite{mstw2008} for signal and background
generation. The corresponding PDF systematics are presented in
Table~III.
\begin{table}[h]
  \label{tab:pdf8}
\begin{center}
\begin{tabular}{||r|cc|cc||r|cc|cc||}\hline
      $Z^*$ mass &  \multicolumn{2}{|c|}{signal} &
      \multicolumn{2}{|c||}{background} &$Z^*$ mass &
      \multicolumn{2}{|c|}{signal} & \multicolumn{2}{|c||}{background} \\
      $$[GeV]   &   $\sigma Br$ [fb] & $\Delta\sigma/\sigma$ [\%]
      & $\sigma Br$ [fb] & $\Delta\sigma/\sigma$ [\%] &
      $$[TeV] & $\sigma Br$ [ab] & $\Delta\sigma/\sigma$ [\%]
      & $\sigma Br$ [ab] & $\Delta\sigma/\sigma$ [\%] \\ \hline
       250  &  75011.  & 3.1 & 408.12 & 2.5 & 2.25 & 1341.4 & 16.6 & 14.809 & 13.7\\
       500  &  5657.3   & 4.2 & 30.944 & 3.4 & 2.50 & 492.04 & 19.8 & 6.4583 & 16.3\\
       750   &  1044.4   & 5.5 & 6.0493 & 4.3 & 2.75 & 179.93 & 23.5 & 2.8387 & 19.5\\
       1000  &  272.90   & 6.9 & 1.6845 & 5.3 & 3.00 & 65.599 & 28.2 & 1.2481 & 22.8\\
       1250  &  84.302   & 8.7 & 0.56248 & 6.5 & 3.25 & 23.814 & 33.1 & 0.54635 & 26.2\\
       1500  &  28.493   & 10.3 & 0.20889 & 7.8 & 3.50 & 8.6385 & 37.4 & 0.23602 & 29.7\\
       1750  &  10.090   & 12.0 & 0.083116 & 9.4 & 3.75 & 3.1289 & 40.6 & 0.10036 & 33.3\\
       2000  &  3.6597   & 14.2 & 0.034575 & 11.3 & 4.00 & 1.1286 & 42.0 & 0.041827 & 36.8\\ \hline
\end{tabular}  \caption{ $Z^\star$ signal and the Standard Model DY background
cross sections times branching fraction in $[M-3\Gamma,M+3\Gamma]$
on-peak region and relative uncertainty due to PDF variation (at
90\% C.L.) with MSTW2008lo90cl set.}
\end{center}
\end{table}

\begin{figure}[h!]\centering
\epsfig{file=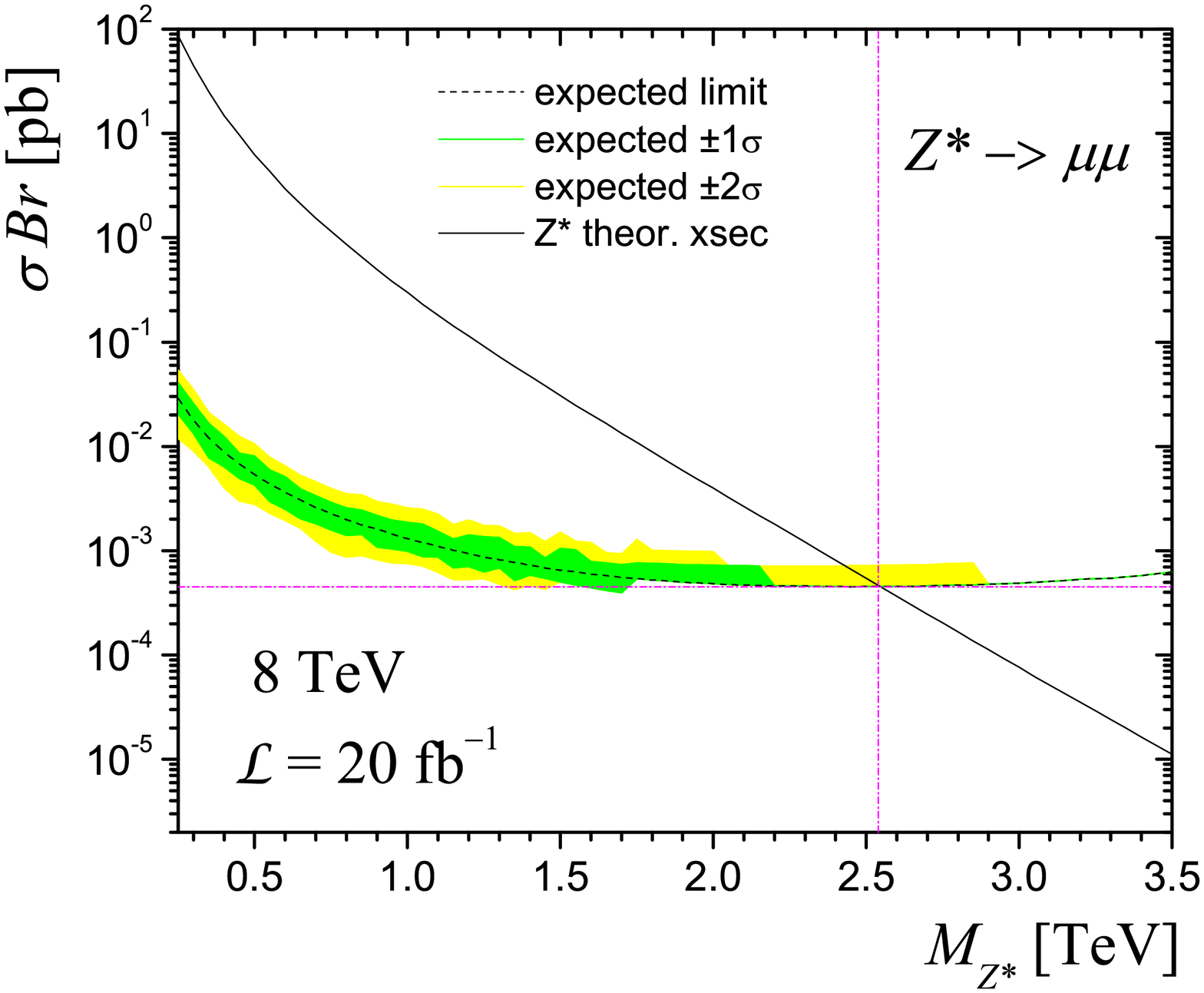,width=0.49\textwidth}
\epsfig{file=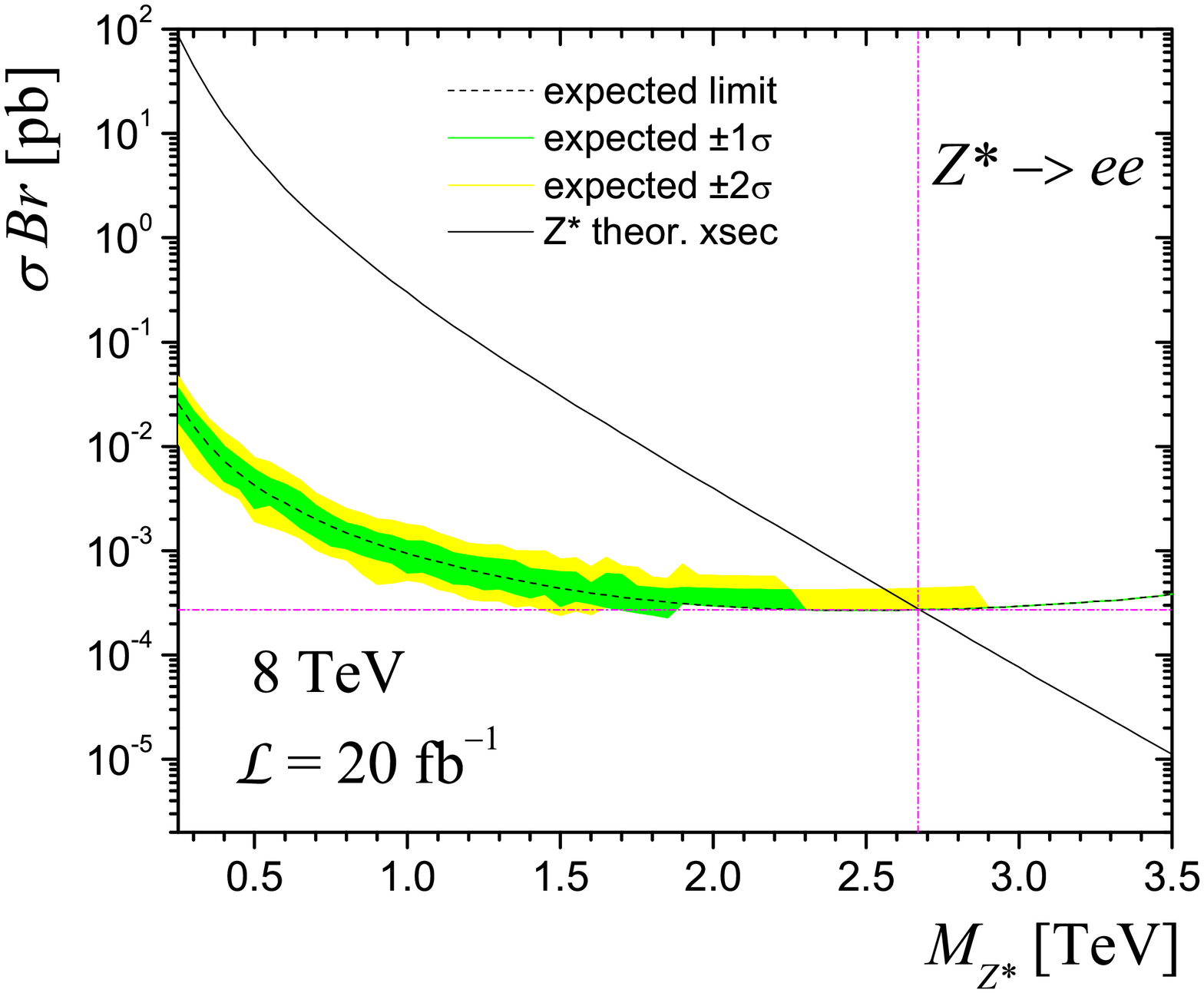,width=0.49\textwidth}
\caption{\label{fig:limit8em} Exclusion limits from muon (left) and
electron (right) channels with 20 fb$^{-1}$ of integrated luminosity
at $\sqrt{s}=8$~TeV.}
\end{figure}

\begin{figure}[h!]\centering
\epsfig{file=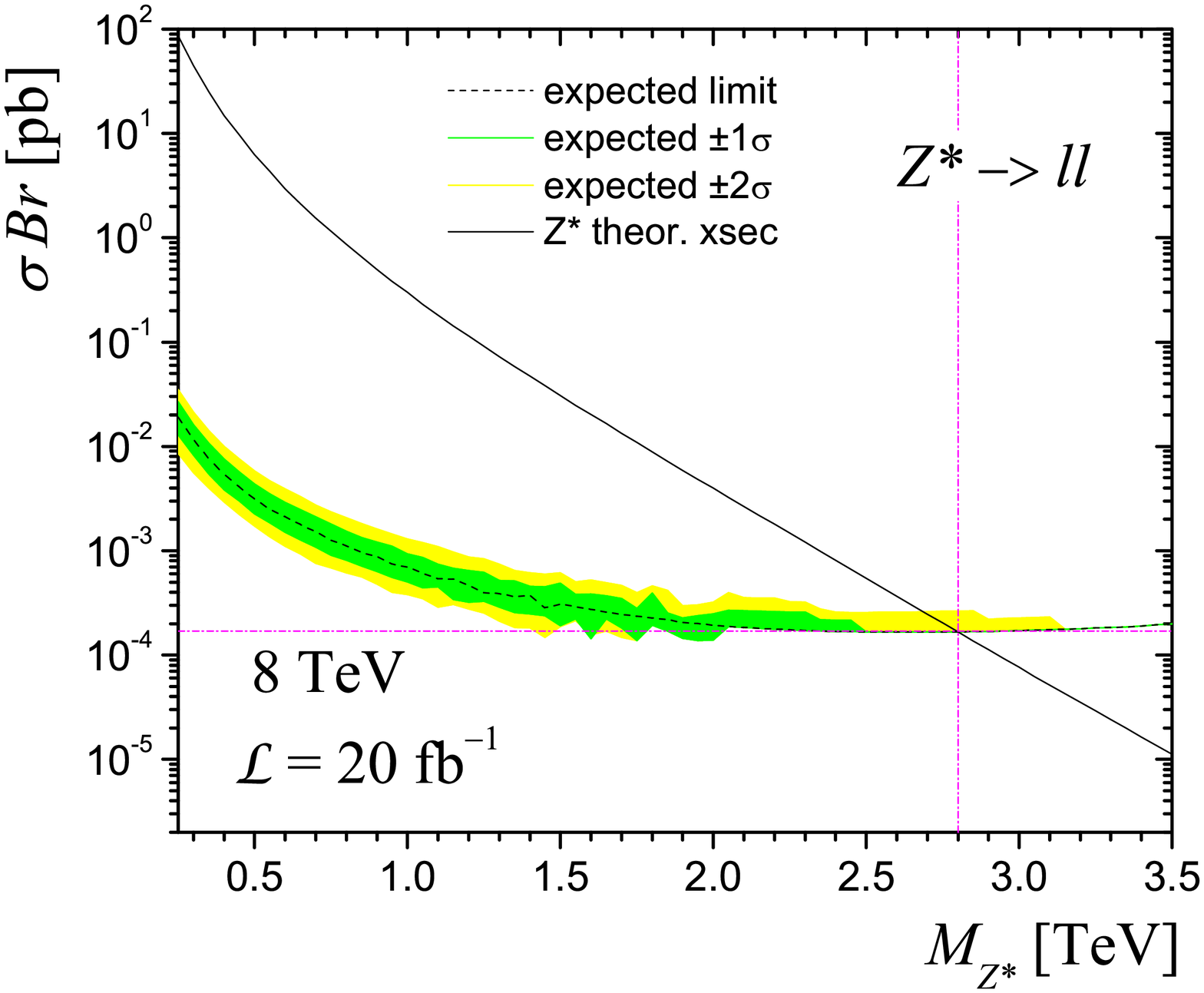,width=0.49\textwidth}
\caption{\label{fig:limit8} Combined exclusion limits with 20
fb$^{-1}$ of integrated luminosity at $\sqrt{s}=8$~TeV.}
\end{figure}

Figure \ref{fig:limit8em} shows the 95\% C.L. expected exclusion
limits on $\sigma Br$ for the electron and muon channels.
The combined limit is shown in Fig.~\ref{fig:limit8}.
Table IV summarizes the constraints on the resonance mass and
$\sigma Br$ of ${\color{red}Z^\star}$ boson.
\begin{table}[h]
  \label{tab:limits8}
\begin{center}
\begin{tabular}{|c|ccc|}
  \hline
   Expected limit & $Z^\star\to \mu^+\mu^-$ & $Z^\star\to e^+e^-$ &
   $Z^\star\to \ell^+\ell^-$ \\
  \hline
  $M$ [TeV] & 2.55 & 2.67 & 2.80 \\
  $\sigma Br$ [fb] & 0.45 & 0.27 & 0.17 \\
  \hline
\end{tabular}
  \caption{The expected 95\% CL limits on the mass and the cross section
  times branching fraction of the $Z^\star$ boson for the $\mu^+\mu^-$ and $e^+e^-$
  channels separately and for their combination at $\sqrt{s}=8$~TeV.}
\end{center}
\end{table}

\section{The discovery potential and the exclusion limits at
$\sqrt{s}=13$~TeV and $\sqrt{s}=14$~TeV.}

In 2015 the LHC will increase the center-of-mass energy up to the
design value. In this section we will estimate the discovery
potential and the exclusion limit, if no deviation from the Standard
Model will be observed. For discovery potential estimation we use
eq.~(\ref{Scl}) with $S_{CL}=5$. At the first stage the
center-of-mass energy could be 13~TeV. Therefore, we will consider
first this possibility.

The corresponding cross sections and PDF systematic uncertainties
are presented in Table~V.
\begin{table}[h]
  \label{tab:pdf13}
\begin{center}
\begin{tabular}{||r|cc|cc||r|cc|cc||}\hline
      $Z^*$ mass &  \multicolumn{2}{|c|}{signal} &
      \multicolumn{2}{|c||}{background} &$Z^*$ mass &
      \multicolumn{2}{|c|}{signal} & \multicolumn{2}{|c||}{background} \\
      $$[TeV]   &   $\sigma Br$ [ab] & $\Delta\sigma/\sigma$ [\%]
      & $\sigma Br$ [ab] & $\Delta\sigma/\sigma$ [\%] &
      $$[TeV] & $\sigma Br$ [zb] & $\Delta\sigma/\sigma$ [\%]
      & $\sigma Br$ [zb] & $\Delta\sigma/\sigma$ [\%] \\ \hline
       2.0  &  32645.   & 8.5 & 217.26 & 6.4 & 4.5 & 56554. & 25.4 & 914.46 & 19.9\\
       2.5  &  8500.8   & 10.6 & 63.858 & 8.1 & 5.0 & 16214. & 32.5 & 330.30 & 24.0\\
       3.0  &  2369.6   & 12.9 & 20.720 & 10.2 & 5.5 & 4635.1 & 40.6 & 117.88 & 28.3\\
       3.5  &  679.60   & 15.8 & 7.1252 & 12.8 & 6.0 & 1321.1 & 49.2 & 41.176 & 32.6\\
       4.0  &  196.25   & 19.9 & 2.5343 & 16.1 & 6.5 & 375.23 & 57.8 & 13.968 & 36.9\\ \hline
\end{tabular}  \caption{ $Z^\star$ signal and the Standard Model DY background
cross sections times branching fraction in $[M-3\Gamma,M+3\Gamma]$
on-peak region and relative uncertainty due to PDF variation (at
90\% C.L.) with MSTW2008lo90cl set at $\sqrt{s}=13$~TeV.}
\end{center}
\end{table}

\begin{figure}[h!]\centering
\epsfig{file=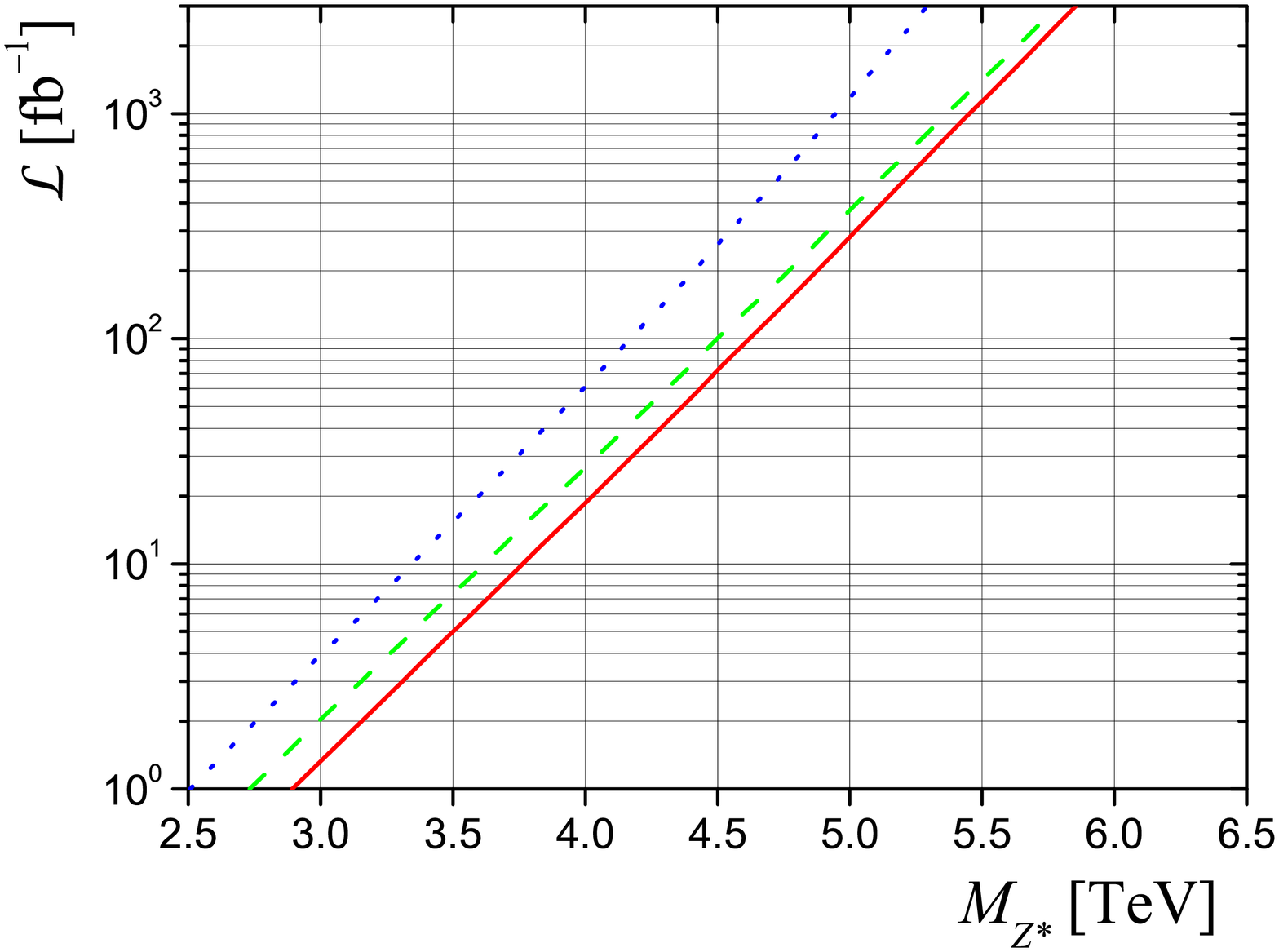,width=0.49\textwidth}
\epsfig{file=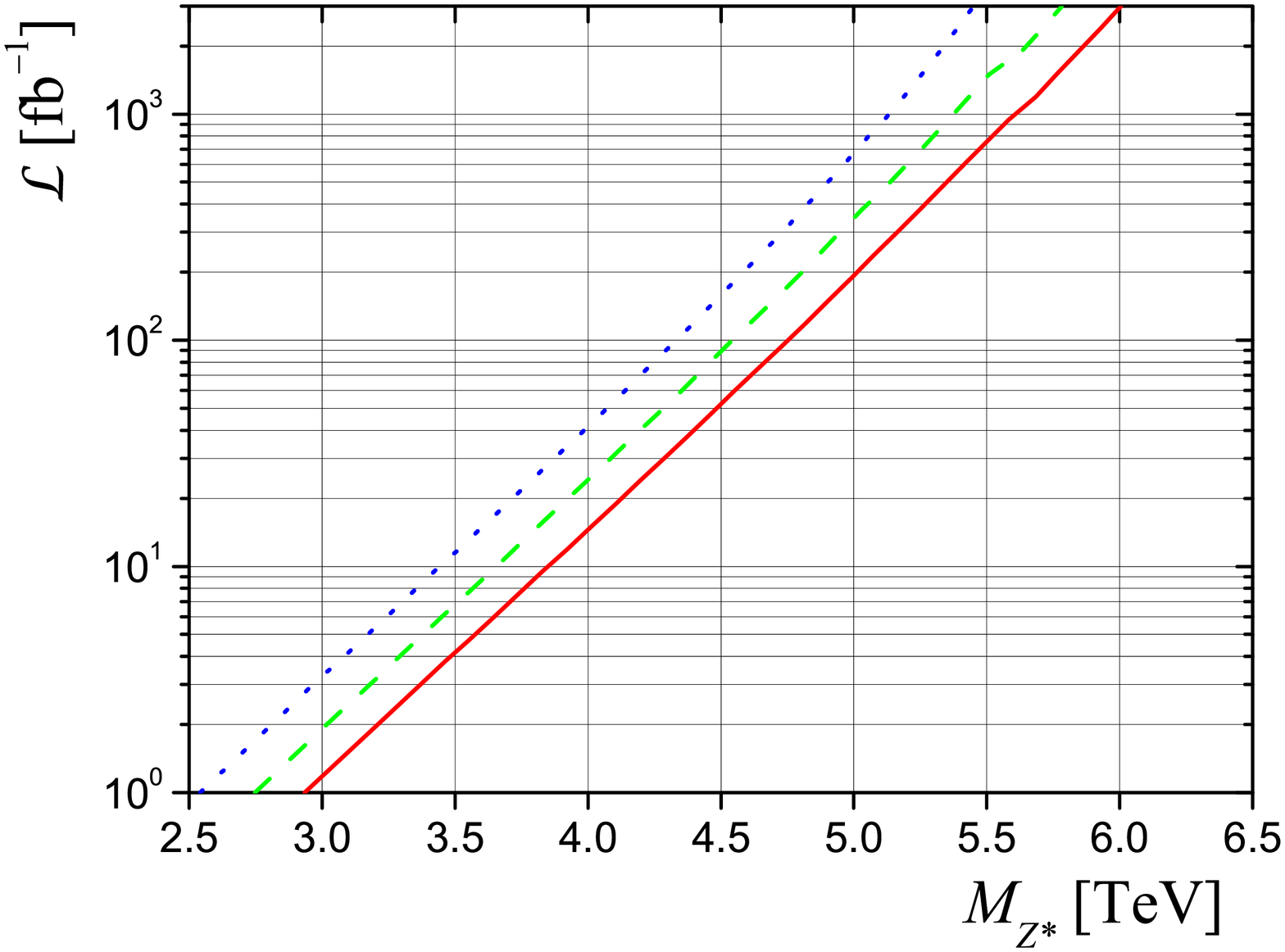,width=0.49\textwidth}
\caption{\label{fig:13} Discovery potential (left) and expected 95\%
CL exclusion limits (right) at $\sqrt{s}=13$~TeV from muon (dotted),
electron (dashed) and combined (solid) channels.}
\end{figure}

Depending on the luminosity they give the following results shown in
Fig.~\ref{fig:13}.
It is interesting to note that higher center-of-mass energy at Run~2
allows to probe higher resonance masses than at Run~1 already with
1~fb$^{-1}$ of an integrated luminosity.

The designed $\sqrt{s}=14$~TeV requires new templates and
systematics (see Table~VI), which only slightly deviates from the
previous case.
\begin{table}[h]
  \label{tab:pdf14}
\begin{center}
\begin{tabular}{||r|cc|cc||r|cc|cc||}\hline
      $Z^*$ mass &  \multicolumn{2}{|c|}{signal} &
      \multicolumn{2}{|c||}{background} &$Z^*$ mass &
      \multicolumn{2}{|c|}{signal} & \multicolumn{2}{|c||}{background} \\
      $$[TeV]   &   $\sigma Br$ [ab] & $\Delta\sigma/\sigma$ [\%]
      & $\sigma Br$ [ab] & $\Delta\sigma/\sigma$ [\%] &
      $$[TeV] & $\sigma Br$ [zb] & $\Delta\sigma/\sigma$ [\%]
      & $\sigma Br$ [zb] & $\Delta\sigma/\sigma$ [\%] \\ \hline
       2.5  &  11603.   & 9.8 & 83.291 & 7.5 & 5.0 & 33566. & 25.8 & 580.80 & 21.1\\
       3.0  &  3471.3   & 11.8 & 28.383 & 9.3 & 5.5 & 10502. & 31.2 & 225.02 & 24.9\\
       3.5  &  1.0776   & 14.1 & 10.295 & 11.5 & 6.0 & 3279.4 & 36.4 & 86.134 & 28.9\\
       4.0  &  339.13   & 17.1 & 3.8814 & 14.2 & 6.5 & 1022.9 & 40.1 & 32.344 & 32.9\\
       4.5  &  106.88   & 20.9 & 1.4963 & 17.4 & 7.0 & 317.37 & 41.5 & 11.820 & 36.9\\ \hline
\end{tabular}  \caption{ $Z^\star$ signal and the Standard Model DY background
cross sections times branching fraction in $[M-3\Gamma,M+3\Gamma]$
on-peak region and relative uncertainty due to PDF variation (at
90\% C.L.) with MSTW2008lo90cl set at $\sqrt{s}=14$~TeV.}
\end{center}
\end{table}

\begin{figure}[h!]\centering
\epsfig{file=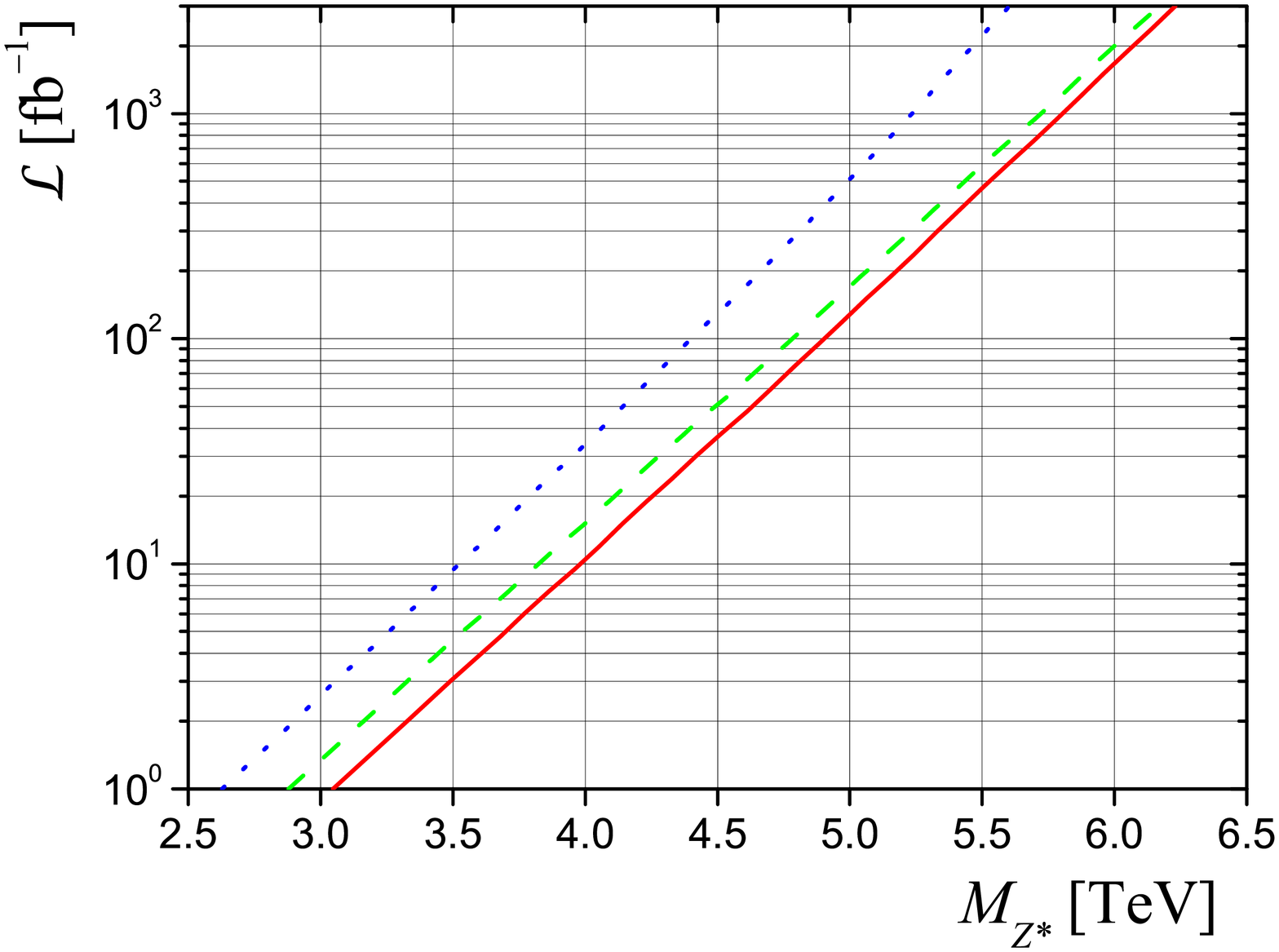,width=0.49\textwidth}
\epsfig{file=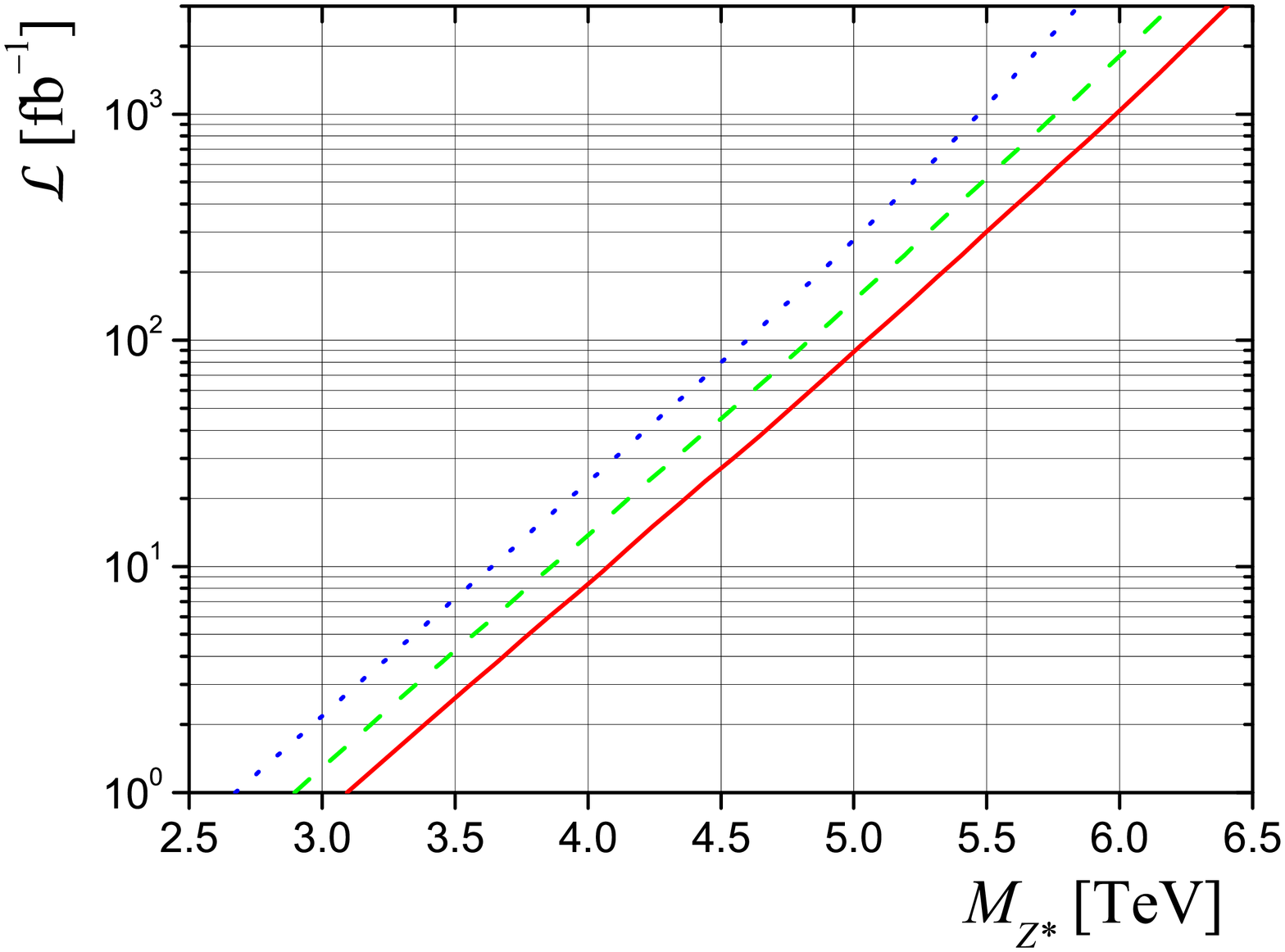,width=0.49\textwidth}
\caption{\label{fig:14} The discovery potential (left) and the
expected 95\% CL exclusion limits (right) at $\sqrt{s}=14$~TeV from
muon (dotted), electron (dashed) and combined (solid) channels.}
\end{figure}

The final results are presented in Fig.~\ref{fig:14}.
In particular, LHC Run~2 can discover $Z^\star$ up to about 5.3~TeV
or exclude it in case of signal absence up to resonance masses of
5.5~TeV. The High Luminosity LHC with 3000~fb$^{-1}$ of an
integrated luminosity can extend that reach in case of a signal to
about 6.2~TeV for discovery or exclude it up to about 6.4~TeV.

\section{The discovery potential and the exclusion limits at $\sqrt{s}=33$~TeV.}

At the end of the paper we will investigate also the discovery
potential and the exclusion limits for the excited boson search in
the case of the highest center-of-mass energy, $\sqrt{s}=33$~TeV.

The corresponding data are given in Table~VII.
\begin{table}[h]
  \label{tab:pdf33}
\begin{center}
\begin{tabular}{||r|cc|cc||r|cc|cc||}\hline
      $Z^*$ mass &  \multicolumn{2}{|c|}{signal} &
      \multicolumn{2}{|c||}{background} &$Z^*$ mass &
      \multicolumn{2}{|c|}{signal} & \multicolumn{2}{|c||}{background} \\
      $$[TeV]   &   $\sigma Br$ [ab] & $\Delta\sigma/\sigma$ [\%]
      & $\sigma Br$ [ab] & $\Delta\sigma/\sigma$ [\%] &
      $$[TeV] & $\sigma Br$ [zb] & $\Delta\sigma/\sigma$ [\%]
      & $\sigma Br$ [zb] & $\Delta\sigma/\sigma$ [\%] \\ \hline
       4.5  &  8741.9   & 7.7 & 56.12 & 5.7 & 9.5 & 48328. & 17.5 & 570.54 & 14.6\\
       5.5  &  2824.5   & 9.3 & 19.769 & 7.0 & 10.5 & 17979. & 20.9 & 251.65 & 17.4\\
       6.5  &  979.46   & 10.9 & 7.6047 & 8.4 & 11.5 & 6681.0 & 24.9 & 111.70 & 20.4\\
       7.5  &  353.68   & 12.7 & 3.0997 & 10.1 & 12.5 & 2476.7 & 29.4 & 49.630 & 23.7\\
       8.5  &  130.17   & 14.9 & 1.3131 & 12.1 & 13.5 & 915.92 & 33.9 & 21.899 & 27.0\\ \hline
\end{tabular}  \caption{ $Z^\star$ signal and the Standard Model DY background
cross sections times branching fraction in $[M-3\Gamma,M+3\Gamma]$
on-peak region and relative uncertainty due to PDF variation (at
90\% C.L.) with MSTW2008lo90cl set at $\sqrt{s}=33$~TeV.}
\end{center}
\end{table}

\begin{figure}[h!]\centering
\epsfig{file=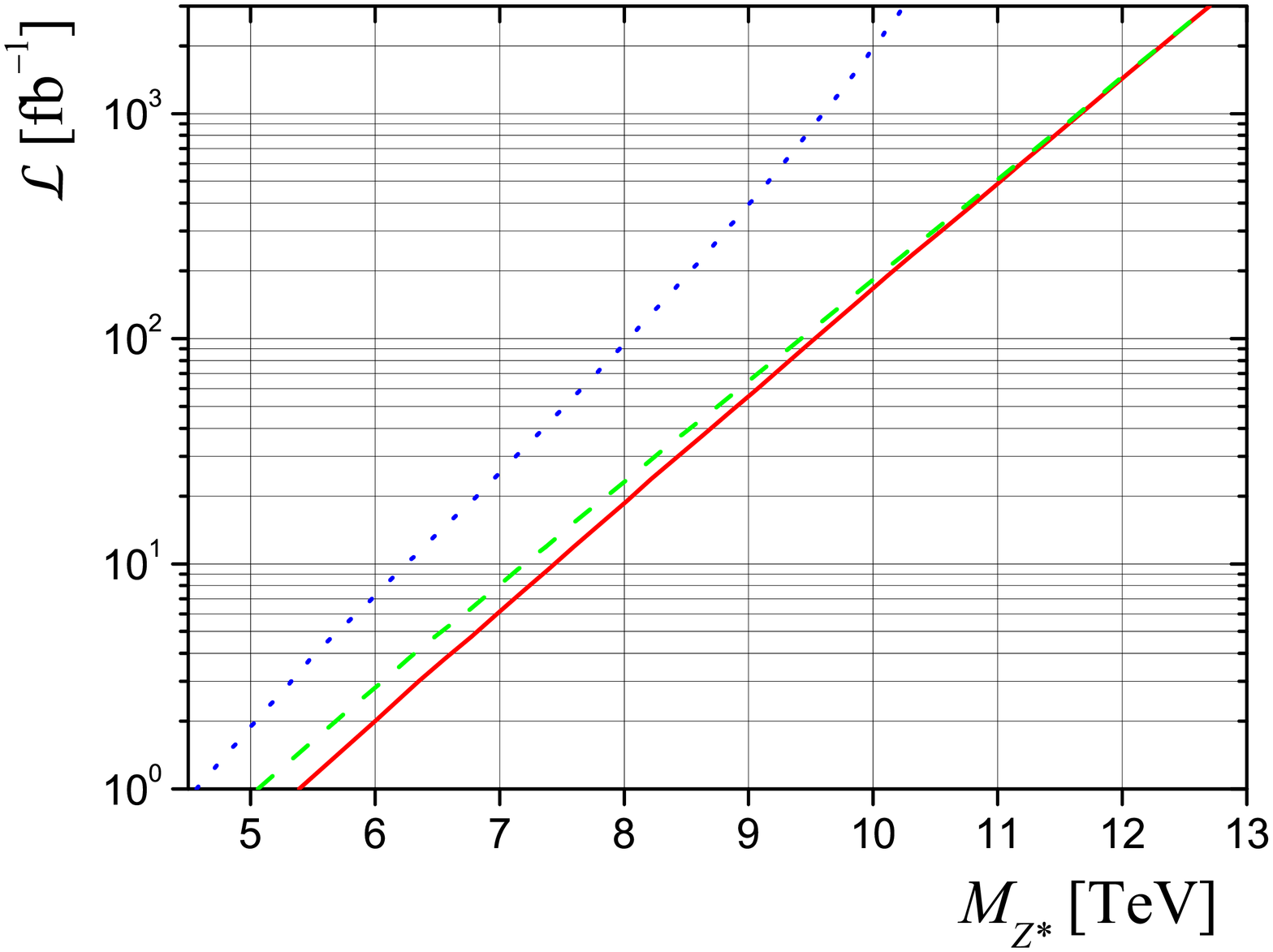,width=0.49\textwidth}
\epsfig{file=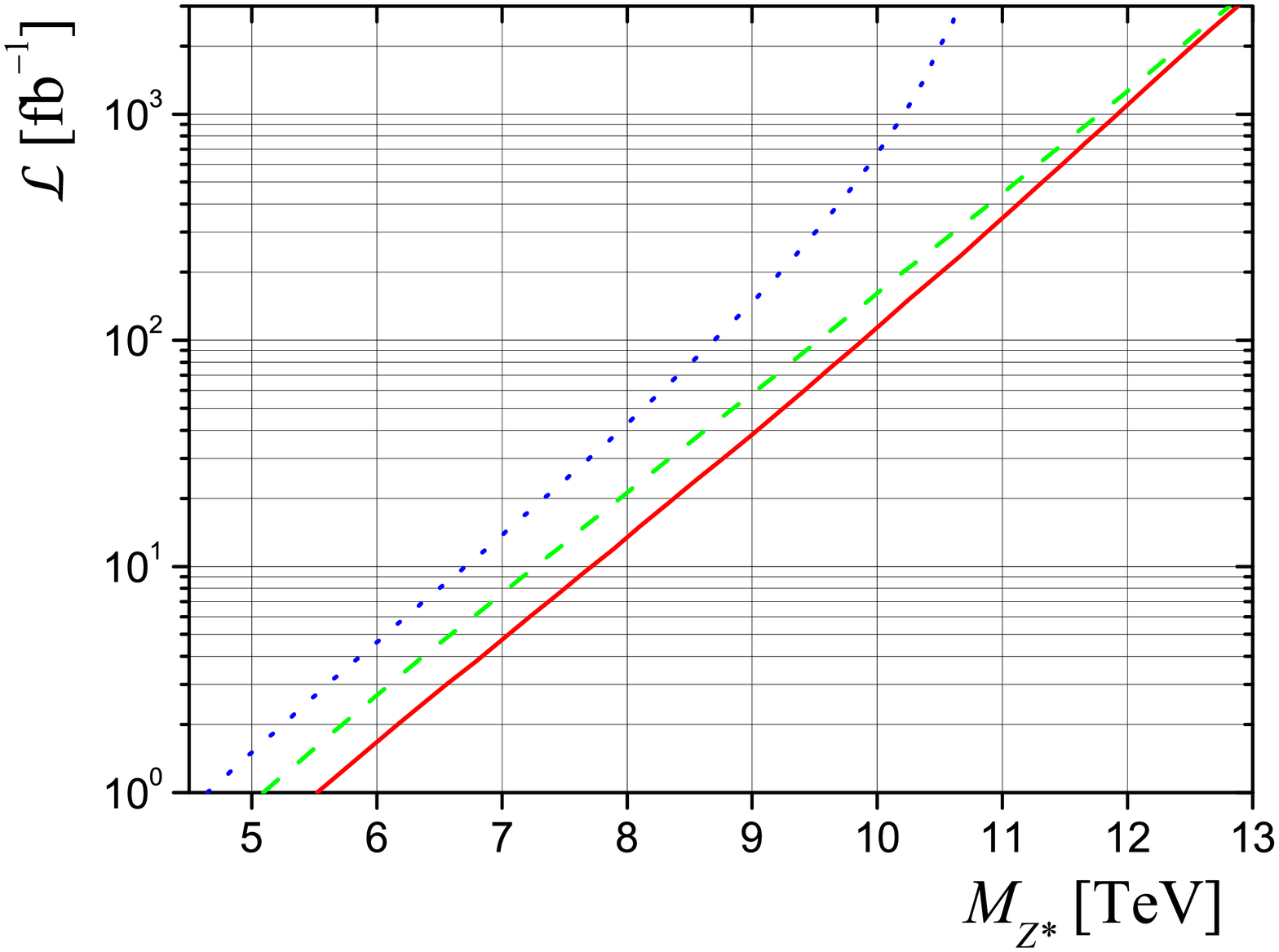,width=0.49\textwidth}
\caption{\label{fig:33} The discovery potential (left) and the
expected 95\% CL exclusion limits (right) at $\sqrt{s}=33$~TeV from
muon (dotted), electron (dashed) and combined (solid) channels.}
\end{figure}

The discovery potential and the exclusion limits on the excited
boson resonance mass depending on the integrated luminosity are
presented in Fig.~\ref{fig:33}.
The plots show that High Energy LHC can probe around two times
heavier resonance masses at the same integrated luminosities than at
Run~2.

\section{Conclusion}
In this paper we have considered the discovery potential and the
exclusion limits on the excited boson search in $pp$ collisions at
the LHC for the different center-of-mass energies and different
luminosities. In particular, LHC Run 2 can discover $Z^\star$ up to
about 5.3 TeV, while the High Luminosity (HL) LHC can extend that
reach to about 6.2 TeV. The High Energy (HE) LHC can probe around
two times heavier resonance masses at the same integrated
luminosities. The field for $Z^\star$ search remains opened both at
the HL-LHC and the HE-LHC.


\begin{thebibliography}{99}
\bibitem{LEP} R. Barate et al. (ALEPH Collaboration), Eur. Phys. J. C
4, 571 (1998); G. Abbiendi et al. (OPAL Collaboration), Phys. Lett.
B 544, 57 (2002); P. Achard et al. (L3 Collaboration), Phys. Lett. B
568, 23 (2003); J. Abdallah et al. (DELPHI Collaboration), Eur.
Phys. J. C 46, 277 (2006).
\bibitem{HERA} C. Adloff et al. (H1 Collaboration), Phys. Lett. B 548, 35
(2002); S.~Chekanov et al. (ZEUS Collaboration), Phys. Lett. B 549,
32 (2002).
\bibitem{Tevatron} D. Acosta et al. (CDF Collaboration), Phys. Rev. Lett.
94, 101802 (2005); V. M. Abazov et al. (D0 Collaboration)
arXiv:0801.0877 [hep-ex].
\bibitem{LHCleptons} ATLAS Collaboration, NJP 15 (2013) 093011,
arXiv:1308.1364 [hep-ex];
CMS Collaboration, Phys. Lett. B 720 (2013) 309, arXiv:1210.2422
[hep-ex].
\bibitem{LHCdijet} ATLAS Collaboration, Phys. Lett. B 708 (2012) 37,
arXiv:1108.6311 [hep-ex];
CMS Collaboration, Phys. Rev. D 87 (2013) 114015, arXiv:1302.4794
[hep-ex].
\bibitem{project} M.V. Chizhov, V.A. Bednyakov, J.A. Budagov, Phys. Atom.
Nucl. 71 (2008) 2096-2100, arXiv:0801.4235 [hep-ph].
\bibitem{doublets} M.V. Chizhov, Mod. Phys. Lett. A 8 (1993) 2753-2762,
hep-ph/0401217; hep-ph/0609141.
\bibitem{CD} M.V. Chizhov, Gia Dvali, Phys. Lett. B703 (2011)
593-598, arXiv:0908.0924 [hep-ph].
\bibitem{ref} M.V. Chizhov, Phys. Part. Nucl. Lett. 8 (2011) 512-516,
arXiv:1005.4287.
\bibitem{Zstar} ATLAS Collaboration, JHEP 11 (2012) 138,
arXiv:1209.2535 [hep-ex].
\bibitem{Wstar} ATLAS Collaboration, Eur. Phys. J. C72 (2012) 2241,
arXiv:1209.4446 [hep-ex].
\bibitem{dilepton} M.V. Chizhov, V.A. Bednyakov, J.A. Budagov,
Phys. Part. Nucl. Lett. 10 (2013) 144-146, arXiv:1109.6876 [hep-ph].
\bibitem{dijet} M.V. Chizhov, V.A. Bednyakov, J.A. Budagov,
Phys. Atom. Nuclei 75 (2012) 90; arXiv:1106.4161 [hep-ph].
\bibitem{CalcHEP} A. Belyaev, N.D. Christensen, A. Pukhov,
Comput. Phys. Commun. 184 (2013) 1729-1769, arXiv:1207.6082
[hep-ph].
\bibitem{hepmdb}
 M. Bondarenko, A. Belyaev, L. Basso, E. Boos, V. Bunichev {\it et al.},
 ``High Energy Physics Model Database : Towards
 decoding of the underlying theory" (within ``Les Houches 2011: Physics
 at TeV Colliders New Physics Working Group Report"), arXiv:1203.1488
 [hep-ph],
 {https://hepmdb.soton.ac.uk}
\bibitem{template} C.P. Hays, A.V. Kotwal, O. Stelzer-Chilton,
Mod. Phys. Lett. A24 (2009) 2387-2403, arXiv:0910.1770 [hep-ex].
\bibitem{CMS} G.L. Bayatian et al., (CMS Collaboration), J. Phys. G34 (2007) 995.
\bibitem{Lathule} M.V. Chizhov, V.A. Bednyakov, J.A. Budagov, Nuovo
Cim. C33 (2010) 343-350, arXiv:1005.2728 [hep-ph].
\bibitem{Zstar10} ATLAS Collaboration, Phys. Lett. B700 (2011)
163-180, arXiv:1103.6218 [hep-ex].
\bibitem{Bayes} A. Caldwell, D. Kollar, K. Kroninger,
Comput. Phys. Commun. 180 (2009) 2197, arXiv:0808.2552
[physics.data-an].
\bibitem{Junk} T. Junk, Nucl. Instrum. Meth. A434 (1999) 435-443, hep-ex/9902006.
\bibitem{cteq6l1} J. Pumplin et al., JHEP 07 (2002) 012, hep-ph/0201195.
\bibitem{cteq61} D. Stump et al., JHEP 10 (2003) 046, hep-ph/0303013.
\bibitem{ATLAS8} ATLAS Collaboration, ATLAS-CONF-2013-017.
\bibitem{mstw2008} A.D. Martin et al., Eur. Phys. J. C63 (2009)
189-285, arXiv:0901.0002 [hep-ph].
\end{thebibliography}
\end{document}